\PassOptionsToPackage{unicode}{hyperref}
\PassOptionsToPackage{hyphens}{url}
\documentclass[
]{article}
\usepackage{amsmath,amssymb}
\usepackage{lmodern}
\usepackage{iftex}
\ifPDFTeX
  \usepackage[T1]{fontenc}
  \usepackage[utf8]{inputenc}
  \usepackage{textcomp} % provide euro and other symbols
\else % if luatex or xetex
  \usepackage{unicode-math}
  \defaultfontfeatures{Scale=MatchLowercase}
  \defaultfontfeatures[\rmfamily]{Ligatures=TeX,Scale=1}
\fi
\IfFileExists{upquote.sty}{\usepackage{upquote}}{}
\IfFileExists{microtype.sty}{% use microtype if available
  \usepackage[]{microtype}
  \UseMicrotypeSet[protrusion]{basicmath} % disable protrusion for tt fonts
}{}
\makeatletter
\@ifundefined{KOMAClassName}{% if non-KOMA class
  \IfFileExists{parskip.sty}{%
    \usepackage{parskip}
  }{% else
    \setlength{\parindent}{0pt}
    \setlength{\parskip}{6pt plus 2pt minus 1pt}}
}{% if KOMA class
  \KOMAoptions{parskip=half}}
\makeatother
\usepackage{xcolor}
\ifLuaTeX
  \usepackage{selnolig}  % disable illegal ligatures
\fi
\IfFileExists{bookmark.sty}{\usepackage{bookmark}}{\usepackage{hyperref}}
\IfFileExists{xurl.sty}{\usepackage{xurl}}{} % add URL line breaks if available
\hypersetup{
  pdftitle={The Most Important Laboratory for Social Scientific and Computing Research in History},
  pdfauthor={Benjamin Mako Hill; Aaron Shaw},
  hidelinks,
  pdfcreator={LaTeX via pandoc}}

\title{The Most Important Laboratory for Social Scientific and Computing
Research in History}
\author{Benjamin Mako Hill \and Aaron Shaw}
\date{October 24, 2019\footnotemark[1]}

\begin{document}
\maketitle

\begingroup
\renewcommand\thefootnote{\fnsymbol{footnote}}
\footnotetext[1]{Published as: Hill, Benjamin Mako, and Aaron Shaw. 2020. “The Most Important Laboratory for Social Scientific and Computing Research in History.” In \textit{Wikipedia @ 20: Stories of an Incomplete Revolution}, edited by Joseph M. Jr. Reagle and Jackie L. Koerner, 159–74. Cambridge, Massachusetts: MIT Press. }
\endgroup

Twenty years ago, Wikipedia's founders could not have dreamed they were
creating the most important laboratory for social scientific and
computing research in history. And yet, that is exactly what has
happened. Wikipedia and its sister projects have launched a thriving
scholarly literature. How thriving? Results from Google Scholar suggest
that over 6,000 scholarly publications mention Wikipedia in their title
and over 1,700,000 mention it somewhere in their text. For comparison,
the phrase ``Catholic Church''---an organization with a nearly 2,000
year head start---returns about the same number of mentions in
publication titles. In under twenty years, Wikipedia has become one of
the most heavily studied organizations of any kind. To the extent that
Wikipedia research is a field of study, what major areas of
investigation have been pursued in the field so far? What are the big
discoveries? The most striking gaps? This essay addresses these
questions and considers some of the most important directions Wikipedia
research might take in the future.

\hypertarget{the-state-of-wikimedia-research}{%
\subsection{The State of Wikimedia
Research}\label{the-state-of-wikimedia-research}}

In 2008, Mako Hill was about to start his first year as a social science
graduate student at MIT where he hoped to study, among other things,
organizational processes that had driven Wikipedia's success. Mako felt
it would behoove him to become better connected to the recent academic
scholarship on Wikipedia. He was also looking for a topic for a talk he
could give at Wikipedia's annual community conference, called
``Wikimania,'' which was going to be hosted by the Library of Alexandria
in Egypt. Attempting to solve both problems at once, Mako submitted a
session proposal for Wikimania suggesting that he would summarize all of
the academic research about Wikipedia published in the previous year in
a talk entitled ``The State of Wikimedia Scholarship 2007--2008.''

Happily, the proposal was accepted. Two weeks before Wikimania, Mako did
a Google Scholar search to build a list of papers he needed to review.
He found himself facing nearly 800 publications. When Mako tried to
import the papers from the search results into his bibliographic
management software, Google Scholar's bot detection software banned his
laptop. Presumably, no human could (or should!) read that many papers.

Mako never did read all the papers that year, but he managed to create a
talk synthesizing some key themes from the previous year in research.
Since then, Mako recruited Aaron to help create new versions of the talk
on a yearly basis. Working together since 2008, the two of us have
collaborated on a ``State of Wikimedia Scholarship'' talk nearly every
year. With a growing cast of collaborators, we sort through the huge
pile of published papers with the term ``Wikipedia'' in their title or
abstracts from the past year. Increasingly, we incorporate papers that
analyze other communities within the
\href{https://wikimedia.org}{Wikimedia projects}. Each time around, we
select 5-8 themes that we think capture major tendencies or innovations
in research published in the previous year. For the presentation, we
summarize each theme and describe an exemplary paper (one per theme) to
the Wikimania audience.

Over the first twenty-years of the project's life, Wikipedia research
has connected researchers who have formed new interdisciplinary field.
We have each coordinated the program of the \emph{International
Symposium on Open Collaboration} (\emph{OpenSym}) a conference started
in 2005 as \emph{WikiSym}. As part of this work, we helped coordinate
papers in a track dedicated to ``Wikipedia and Wikimedia research.''
Each year the \emph{Web Conference} (formerly \emph{WWW}) hosts a
workshop that focuses on Wikipedia and Wikimedia research. Since 2011,
volunteers have helped create a monthly ``Wikimedia Research
Newsletter'' which is published in English Wikipedia's newsletter
\emph{The Signpost} and provides a sort of monthly version of our annual
talk. The Wikimedia Foundation runs a monthly ``research showcase''
where researchers from the around the world can present their work.
There is an active mailing list for Wikimedia researchers.

Figure 1: Number of items returned for Google Scholar for publications
containing ``wikipedia'' in the title by year of publication.

As the graph in Figure 1 suggests, these venues capture only a tiny
fraction of Wikimedia research. Our attempts to characterize this body
of research draw from our experience preparing the annual Wikimania talk
each year as well as from our experience in these other spaces. Like our
Wikimania talk, this chapter remains incomplete and aims to provide a
brief tour of several important themes. Others have published literature
reviews of Wikipedia and Wikimedia research which make attempts to
provide more comprehensive---although still
limited---approaches.\textsuperscript{\href{file:///home/mako/research/wp20_chapters/wp20-wikipedia_research.html\#fn1}{1}}
Our experience watching Wikipedia scholarship grow and shift has led to
one overarching conclusion: Wikipedia has become part of the mainstream
of every social and computational research field we know of. Some areas
of study, such as the analysis of human computer interaction, knowledge
management and information systems, and online communication, have
undergone profound shifts in the past twenty years that have been
partially driven by Wikipedia research. In this process, Wikipedia has
acted as a shared object of study that has connected a range of
disparate academic fields to each other.

\hypertarget{wikipedia-as-a-source-of-data}{%
\subsection{Wikipedia as a Source of
Data}\label{wikipedia-as-a-source-of-data}}

Perhaps the most widespread and pervasive form of Wikipedia research is
not research ``about'' Wikipedia at all, but research that uses
Wikipedia as a convenient dataset to study something else. This was the
only theme that showed up every single year during the nine years that
we presented the ``State of Research'' review.

In 2017, Mohamed Medhdi and a team published a systematic literature
review of 132 papers that use Wikipedia as a ``corpus'' of
human-generated
text.\textsuperscript{\href{file:///home/mako/research/wp20_chapters/wp20-wikipedia_research.html\#fn2}{2}}
Most of these papers come from the engineering field of information
retrieval (IR) where the goal is to devise approaches for calling up
particular information from a database. Wikipedia opens a wide range of
tasks in IR research because it provides a vast database of useful
knowledge that is tagged with categories and metadata---but not in the
typically ``structured'' way required by databases.

Another large group of examples come from the field of natural language
processing (NLP) which exists at the intersection of computer science
and engineering and linguistics. NLP research designs and evaluates
approaches for parsing, understanding, and sometimes generating
human-intelligible language. As with IR, Wikipedia presents an
opportunity to NLP research because it encompasses an enormous,
multilingual dataset written and categorized by humans about a wide
variety of topics. Wikipedia has proven invaluable as a dataset for
these applications because it is ``natural'' in the sense that humans
wrote it, because it is made freely available in ways that facilitate
computational analysis, and because it exists in hundreds of languages.
Nearly half of the papers in Mehdi's review study a version of Wikipedia
other than English and more than a third of the papers look at more than
one language edition Wikipedia.

Recently, Wikipedia has spawned a large number of ``derivative''
datasets and databases that extract data from Wikipedia for studying a
wide variety of topics. Similarly, a large body of academic research has
focused on building tools to transform data from Wikipedia and extract
specific subsets of data. One of the newest Wikimedia projects,
Wikidata, extends these benefits by creating a new layer of structured
data that is collaboratively authored and edited like Wikipedia but that
formally represents underlying relationships between entities that may
be the topics of Wikipedia articles. As Wikipedia and Wikidata continue
to grow and render ideas and language more amenable to computational
processing, their value as a dataset and data source to researchers is
also increasing.

\hypertarget{the-gender-gap}{%
\subsection{The Gender Gap}\label{the-gender-gap}}

In 2008, the results of a large opt-in survey of Wikipedia editors
suggested that upwards of 80\% of editors to Wikipedia across many
language editions were male. The finding sent shockwaves through both
the Wikipedia editor and research communities and was widely reported on
in the press. Both the Wikimedia Foundation and Wikipedia community have
responded by making ``the gender gap'' a major strategic priority and
have poured enormous resources into addressing the disparity. Much of
this work has involved research. As a result, issues related to gender
have been a theme in our report on Wikipedia research nearly every year
since 2012.

One series of papers have aimed to characterize the ``gender gap.'' This
work adopted better sampling methods, adjusted for bias in survey
response, and, in at least one case, commissioned a nationally
representative sample of adults in the United States who were asked
about their Wikipedia contribution
behavior.\textsuperscript{\href{file:///home/mako/research/wp20_chapters/wp20-wikipedia_research.html\#fn3}{3}}
Some recent projects have also begun to unpack the ``gap'' by looking at
the ways in which it
emerges.\textsuperscript{\href{file:///home/mako/research/wp20_chapters/wp20-wikipedia_research.html\#fn4}{4}}
Although this follow-on work presented a range of different estimates of
the scope of the gap in participation between male and female editors,
none of the work overturned the basic conclusion that Wikipedia's
editor-base appears largely, if not overwhelmingly, made up of men.

Another group of studies examine different gender gaps including gaps in
content coverage. For example, research has found that women and people
of color are systematically less likely than similarly notable white men
to have
articles.\textsuperscript{\href{file:///home/mako/research/wp20_chapters/wp20-wikipedia_research.html\#fn5}{5}}
Other work has shown that Wikipedia's content tends to suffer a range of
gender biases and gaps as well---for example, by using terms and images
that tends to reflect existing gender
bias.\textsuperscript{\href{file:///home/mako/research/wp20_chapters/wp20-wikipedia_research.html\#fn6}{6}}

Some work has also connected explanations of the gender gap among
contributors to inequality and bias in articles. Existing Wikipedia
communities may deter women and others from editing and may define and
enforce criteria for article creation in ways that differentially impact
articles about or of interest to
women.\textsuperscript{\href{file:///home/mako/research/wp20_chapters/wp20-wikipedia_research.html\#fn7}{7}}

The work on the gender gap in Wikipedia began with a strong focus on
gender inequality within Wikipedia and among Wikipedia editors. More
recent work has sought to understand how Wikipedia content may reflect
underlying inequalities and patterns of stratification in the world in
some other ways. This work has shown that by studying gendered and other
types of inequality in Wikipedia, we can learn about some of the
mechanisms of social stratification more broadly.

\hypertarget{content-quality-and-integrity}{%
\subsection{Content Quality and
Integrity}\label{content-quality-and-integrity}}

Research into content quality and integrity on Wikipedia has also been
an enduring focus of Wikipedia research. In a 2005 piece that is one of
the most widely-discussed examples of Wikipedia research, Jim Giles at
\emph{Nature} ran an informal study distributing a set of Wikipedia and
\emph{Encyclopedia Britannica} articles to experts and asking them to
identify errors in
each.\textsuperscript{\href{file:///home/mako/research/wp20_chapters/wp20-wikipedia_research.html\#fn8}{8}}
The expert coders found about the same number of errors in each group,
leading to the conclusion---surprising at the time---that Wikipedia
articles might be comparable to those produced by professionals and
experts. The early \emph{Nature} study has been reproduced in larger
samples with results that suggest that, over time, Wikipedia typically
surpasses general encyclopedias like
\emph{Britannica}.\textsuperscript{\href{file:///home/mako/research/wp20_chapters/wp20-wikipedia_research.html\#fn9}{9}}
Perhaps more influentially, the template of the Giles study has been
repeated over and over again in various knowledge domains that include
drug information, mental disorders, otolaryngology---just to name
several topics in
medicine.\textsuperscript{\href{file:///home/mako/research/wp20_chapters/wp20-wikipedia_research.html\#fn10}{10}}

Of course, quality itself is much more complicated and multidimensional
than the sum of factual errors in a sample of articles. A number of
studies have tried to assess quality in other terms. Some consider the
relative neutrality of articles on contentious
topics.\textsuperscript{\href{file:///home/mako/research/wp20_chapters/wp20-wikipedia_research.html\#fn11}{11}}
Others look for the absence of important information. Wikipedians
regularly evaluate the quality of their own articles in terms of
comprehensiveness, writing style, the number and reliability of
references, and adherence to Wikipedia's own policies. There have been a
series of attempts to adapt these types of quality measures
qualitatively. This work seems to indicate that although Wikipedia is
enormous, many topics are covered in ways that are
superficial.\textsuperscript{\href{file:///home/mako/research/wp20_chapters/wp20-wikipedia_research.html\#fn12}{12}}
Overall, this body of research has shown the quality of the material
that is covered is high.

Some of the most exciting work on these issues has examined the social
processes that lead to relatively higher or lower article quality. For
example, although quality and viewership of articles are related, a few
recent studies have measured the degree to which topics are
``underproduced'' relative to readers'
interest.\textsuperscript{\href{file:///home/mako/research/wp20_chapters/wp20-wikipedia_research.html\#fn13}{13}}
Another paper shows that articles on contentious topics edited by more
ideologically polarized editors tend to become higher quality than those
with less diverse editor
groups.\textsuperscript{\href{file:///home/mako/research/wp20_chapters/wp20-wikipedia_research.html\#fn14}{14}}
Other work has sought to understand how readers of Wikipedia perceive
quality.\textsuperscript{\href{file:///home/mako/research/wp20_chapters/wp20-wikipedia_research.html\#fn15}{15}}
In an era where factual information is increasingly contested and
polarized, this line of inquiry offers promise of general insights into
the means of producing and sustaining reliable, high quality public
knowledge resources.

\hypertarget{wikipedia-and-education}{%
\subsection{Wikipedia and Education}\label{wikipedia-and-education}}

Early on in its ascendance, many viewed Wikipedia as a threat to
educational authority and a source of dubious information. Initial
research on Wikipedia in education documented the ways that students
used Wikipedia and, in general, suggested that students were relying on
Wikipedia heavily as a first stop for information on a given subject.
For many teachers, Wikipedia's open editing policy made its content
inherently problematic, if not inherently incompatible, with formal
institutions of teaching and learning.

The study of Wikipedia in education has evolved enormously. In part,
educators have changed their attitudes about the site and some studies
have attempted to document these
shifts.\textsuperscript{\href{file:///home/mako/research/wp20_chapters/wp20-wikipedia_research.html\#fn16}{16}}
The focus of academic writing about the pedagogical role of Wikipedia is
no longer on either the question of \emph{if} students use Wikipedia or
how to discourage them from doing so. Instead, researchers of Wikipedia
in education now focus on how to engage students in contributing to
Wikipedia as part of coursework.

Partly, this change seems driven by the success of the Wiki Education
Foundation---a spin-off of the Wikimedia Foundation that supports
instructors of higher education in incorporating Wikipedia into their
classes. Numerous papers and book chapters now document these
experiences. One example from psychology describes the way that 93
students in an introductory human development course helped improve
Wikipedia coverage of basic information on human development on the
web.\textsuperscript{\href{file:///home/mako/research/wp20_chapters/wp20-wikipedia_research.html\#fn17}{17}}

\hypertarget{viewership}{%
\subsection{Viewership}\label{viewership}}

The large majority of research on Wikipedia has focused on its content
and the social systems that produce it. But Wikipedia isn't only an
enormous corpus created by millions, it is also one of the top ten most
popular websites on earth---visited by billions of people each year. In
2007, the Wikimedia Foundation started publishing data that summarized
what visitors to Wikipedia have looked at. This data has now led to a
large body of research on the viewership of the encyclopedia.

Some work on viewership takes advantage of Wikipedia's general
usefulness and uses which pages people visit as an index of how people
allocate their attention. For example, the Snowden revelations led to
chilling effects whereby people became systematically less likely to
look at certain sensitive
topics.\textsuperscript{\href{file:///home/mako/research/wp20_chapters/wp20-wikipedia_research.html\#fn18}{18}}
Other studies have used Wikipedia viewership data to predict the
prevalence of illnesses and influenza, box office revenue, election
results in a number of countries, or simply to capture a
\emph{zeitgeist}.\textsuperscript{\href{file:///home/mako/research/wp20_chapters/wp20-wikipedia_research.html\#fn19}{19}}

Scholars have also combined data on Wikipedia viewership with editing
data to understand the relationship between the consumption and
production of knowledge. Some early work in this area considered whether
viewership related to participation in editing and content
quality.\textsuperscript{\href{file:///home/mako/research/wp20_chapters/wp20-wikipedia_research.html\#fn20}{20}}
Others have tried to model relatively complex dynamics through which
viewers become editors and help produce the
encyclopedia.\textsuperscript{\href{file:///home/mako/research/wp20_chapters/wp20-wikipedia_research.html\#fn21}{21}}

\hypertarget{organization-and-governance}{%
\subsection{Organization and
Governance}\label{organization-and-governance}}

When Wikipedia was first founded, one of the most urgent areas of
inquiry focused on the organization and governance of the project.
Seminal work by Benkler suggested that Wikipedia used technology to
organize knowledge production in transformative ways. Since then,
research on the organization of Wikipedia has grown steadily, often in
an attempt to explain its arguably shocking
success.\textsuperscript{\href{file:///home/mako/research/wp20_chapters/wp20-wikipedia_research.html\#fn22}{22}}

Research has sometimes treated Wikipedia as a community of communities
to investigate collaborative processes. For example, both article-level
collaborations and organized editing efforts in the form of
\emph{WikiProjects} have attracted extensive research. Perhaps not
surprisingly, WikiProjects appear to struggle with many of the same
kinds of organizational challenges that affect collaborative efforts
elsewhere.\textsuperscript{\href{file:///home/mako/research/wp20_chapters/wp20-wikipedia_research.html\#fn23}{23}}
Many studies of organization within Wikipedia have found creative ways
to document and describe otherwise familiar patterns and have sometimes
revealed distinctions between more familiar organizational practices and
those pursued in a large, distributed, online volunteer effort like
Wikipedia.

We have been involved in some related work that challenges the
``stylized facts'' about Wikipedia's organization and which has
suggested some of the ways that Wikipedia's mode of organization and
governance may be
limited.\textsuperscript{\href{file:///home/mako/research/wp20_chapters/wp20-wikipedia_research.html\#fn24}{24}}
We also also advocated for comparative studies that look beyond
Wikipedia---and English Wikipedia in particular---in order to draw more
general understandings of the organizational processes
involved.\textsuperscript{\href{file:///home/mako/research/wp20_chapters/wp20-wikipedia_research.html\#fn25}{25}}
Wikipedia includes hundreds of more-or-less completely distinct language
communities with different experiences and with different degrees of
success. For instance, several of our papers and others' undermine the
widespread perception that Wikipedia's style of organizing does not
entail hierarchies or other patterns of entrenchment among early
community
leaders.\textsuperscript{\href{file:///home/mako/research/wp20_chapters/wp20-wikipedia_research.html\#fn26}{26}}
A small number of studies have engaged in comparative work that studies
Wikipedia across numerous language editions, illustrating the diversity
of collaborative
dynamics.\textsuperscript{\href{file:///home/mako/research/wp20_chapters/wp20-wikipedia_research.html\#fn27}{27}}

As a large population of organizations, Wikipedia offers a data source
of exceptional granularity. Nevertheless, scholars continue to struggle
to understand how Wikipedia is like and unlike more traditional
organizations. We still know little about when the experience of
traditional organizations will be instructive to Wikipedia. For example,
in our own work we found that an attempt to import newcomer
socialization practices with a long history of success in traditional
organizations seemed to have little effect on newcomer retention in
Wikpiedia.\textsuperscript{\href{file:///home/mako/research/wp20_chapters/wp20-wikipedia_research.html\#fn28}{28}}
In a related sense, we still know little about when the things we learn
about organization in Wikipedia will---or will not---translate into
other spaces.

\hypertarget{wikipedia-in-the-world}{%
\subsection{Wikipedia in the World}\label{wikipedia-in-the-world}}

The metaphor of a laboratory we used in our introduction depicts
Wikipedia as somehow isolated from the rest of the world. However,
Wikipedia \emph{affects} the world in important ways as well. Some
exciting studies have investigated specific aspects of this
relationship.

The earliest versions of this work simply documented the ways that
Wikipedia became increasingly integrated into many people's everyday
lives. One striking example from 2009 described the growing rate at
which legal opinion and published law relied on citations to Wikipedia
to establish facts about the world in hundreds of legal opinions in the
US District Courts and Courts of
Appeals.\textsuperscript{\href{file:///home/mako/research/wp20_chapters/wp20-wikipedia_research.html\#fn29}{29}}
Other work looks at how Wikipedia content is increasingly syndicated
into other places and suggests that an enormous portion of all
successful Internet searches would be failures if Wikipedia did not
exist.\textsuperscript{\href{file:///home/mako/research/wp20_chapters/wp20-wikipedia_research.html\#fn30}{30}}

Given its prominence in search engine rankings, a group of
scholars---primarily economists---have come to Wikipedia as a platform
on which to run experiments \emph{on the world}. For example, one group
improved a random set of articles about small European cities and showed
that tourism traffic improved relative to a control group whose articles
were not
improved.\textsuperscript{\href{file:///home/mako/research/wp20_chapters/wp20-wikipedia_research.html\#fn31}{31}}
Another study showed that improving a randomly selected set of Wikipedia
articles about scientific studies tends to increase the citations to the
studies mentioned in articles and tends to shape the language subsequent
research studies use when they describe the cited
work.\textsuperscript{\href{file:///home/mako/research/wp20_chapters/wp20-wikipedia_research.html\#fn32}{32}}

These studies do more than show that Wikipedia is important---although
they certainly do that. They provide important evidence in favor of
particular theories of information diffusion and they document the way
that knowledge is created and spreads. In this way, Wikipedia provides
not only a laboratory for studying social processes, but acts as a key
piece of laboratory equipment for studying social behavior ``in the
wild.''

\hypertarget{conclusion}{%
\subsection{Conclusion}\label{conclusion}}

Insights about how the largest volunteer effort in the world have
managed to produce the largest encyclopedias in history will continue to
advance the frontiers of scientific knowledge. Understanding how
Wikipedia and projects like it work can help us organize other parts of
social life more effectively.

We conclude with an invocation to researchers to think about Wikipedia
even more, and in even broader ways. Wikipedia is the most influential
and widely accessed free information resource on the internet as well as
the most widely used information platform in human history. As such,
Wikipedia merits comparisons to other epochal transformations in how
humans collect, organize, store, and disseminate ideas. It deserves the
scholarly attention it has received. In particular, understanding how
and why communities like Wikipedia manage to mobilize vast numbers of
volunteers and sustain such high quality, large scale, information
resources means looking beyond the boundaries of Wikipedia to conduct
comparisons, impact evaluations, and more. That ought to keep us all
busy for at least another twenty years.

\emph{Acknowledgments: This work was supported by the National Science
Foundation (awards IIS-1617129 and IIS-1617468).}

\begin{center}\rule{0.5\linewidth}{0.5pt}\end{center}

\begin{enumerate}
\item
  Chitu Okoli, Mohamad Mehdi, Mostafa Mesgari, Finn Årup Nielsen, and
  Arto Lanamäki, ``Wikipedia in the Eyes of Its Beholders: A Systematic
  Review of Scholarly Research on Wikipedia Readers and Readership,''
  \emph{Journal of the Association for Information Science and
  Technology} 65, no. 12 (2014): 2381--2403,
  \url{https://doi.org/10.1002/asi.23162}; Mohamad Mehdi, Chitu Okoli,
  Mostafa Mesgari, Finn Årup Nielsen, and Arto Lanamäki. ``Excavating
  the Mother Lode of Human-Generated Text: A Systematic Review of
  Research That Uses the Wikipedia Corpus,'' \emph{Information
  Processing \& Management} 53, no. 2 (March 1, 2017): 505--29,
  \url{https://doi.org/10.1016/j.ipm.2016.07.003}.
\item
  Mohamad Mehdi et al., ``Excavating the Mother Lode of Human-Generated
  Text.''
\item
  Eszter Hargittai and Aaron Shaw, ``Mind the Skills Gap: The Role of
  Internet Know-How and Gender in Differentiated Contributions to
  Wikipedia,'' \emph{Information, Communication \& Society} 18, no. 4
  (April 3, 2015): 424--42,
  \url{https://doi.org/10.1080/1369118X.2014.957711}; Benjamin Mako Hill
  and Aaron Shaw, ``The Wikipedia Gender Gap Revisited: Characterizing
  Survey Response Bias with Propensity Score Estimation,'' \emph{PLoS
  ONE} 8, no. 6 (June 26, 2013): e65782,
  \url{https://doi.org/10.1371/journal.pone.0065782}; Aaron Shaw and
  Eszter Hargittai, ``The Pipeline of Online Participation Inequalities:
  The Case of Wikipedia Editing.'' \emph{Journal of Communication} 68,
  no. 1 (February 1, 2018): 143--68.
  \url{https://doi.org/10.1093/joc/jqx003}.
\item
  e.g., Shaw and Hargittai, ``The Pipeline of Online Participation
  Inequalities.''
\item
  e.g., Joseph Reagle and Lauren Rhue, ``Gender Bias in Wikipedia and
  Britannica,'' \emph{International Journal of Communication} 5 (2011):
  1138--58, \url{http://ijoc.org/index.php/ijoc/article/view/777}; Julia
  Adams, Hannah Brückner, and Cambria Naslund, ``Who Counts as a Notable
  Sociologist on Wikipedia? Gender, Race, and the `Professor Test',''
  \emph{Socius} 5 (January 1, 2019): 2378023118823946,
  \url{https://doi.org/10.1177/2378023118823946}.
\item
  Max Klein and Piotr Konieczny, ``Wikipedia in the World of Global
  Gender Inequality Indices: What the Biography Gender Gap Is
  Measuring,'' in \emph{Proceedings of the 11th International Symposium
  on Open Collaboration}, OpenSym '15 (New York, NY: ACM, 2015),
  16:1--16:2, \url{https://doi.org/10.1145/2788993.2789849}; Claudia
  Wagner, David Garcia, Mohsen Jadidi, and Markus Strohmaier, ``It's a
  Man's Wikipedia? Assessing Gender Inequality in an Online
  Encyclopedia,'' in \emph{Proceedings of the Ninth International AAAI
  Conference on Web and Social Media (ICWSM '15)} (Palo Alto, CA: AAAI,
  2015), 454--63,
  \url{https://www.aaai.org/ocs/index.php/ICWSM/ICWSM15/paper/view/10585};
  Olga Zagovora, Fabian Flöck, and Claudia Wagner, ``\,`(Weitergeleitet
  von Journalistin)': The Endered Presentation of Professions on
  Wikipedia,'' in \emph{Proceedings of the 2017 ACM on Web Science
  Conference (WebSci '17)} (New York, NY: ACM, 2017), 83--92,
  \url{https://doi.org/10.1145/3091478.3091488}.
\item
  Julia B. Bear and Benjamin Collier, ``Where Are the Women in
  Wikipedia? Understanding the Different Psychological Experiences of
  Men and Women in Wikipedia,'' \emph{Sex Roles} 74, nos. 5-6 (January
  4, 2016): 254--65, \url{https://doi.org/10.1007/s11199-015-0573-y};
  Menking, Amanda, Ingrid Erickson, and Wanda Pratt. ``People Who Can
  Take It: How Women Wikipedians Negotiate and Navigate Safety,'' in
  \emph{Proceedings of the 2019 CHI Conference on Human Factors in
  Computing Systems}, 472:1--472:14. CHI '19. New York, NY, USA: ACM,
  2019, \url{https://doi.org/10.1145/3290605.3300702}; Heather Ford and
  Judy Wajcman, ``\,`Anyone Can Edit', Not Everyone Does: Wikipedia's
  Infrastructure and the Gender Gap,'' \emph{Social Studies of Science}
  47, no. 4 (2017): 511--27,
  \url{https://doi.org/10.1177/0306312717692172}.
\item
  Jim Giles, ``Internet Encyclopaedias Go Head to Head,'' \emph{Nature}
  438, no. 7070 (December 14, 2005): 900--901,
  \url{https://doi.org/10.1038/438900a}.
\item
  Fernando Silvério Nifrário Rodrigues, ``Colaboração Em Massa Ou
  Amadorismo Em Massa? Um Estudo Comparativo Da Qualidade Da Informação
  Científica Produzida Utilizando Os Conceitos E Ferramentas Wiki''
  (Universidade de Évora, 2012),
  \url{http://massamateurism.blogspot.co.uk/p/synopsis.html}.
\item
  Kevin A. Clauson, Hyla H. Polen, Maged N. Kamel Boulos, and Joan H.
  Dzenowagis. ``Scope, Completeness, and Accuracy of Drug Information in
  Wikipedia,'' \emph{The Annals of Pharmacotherapy} 42, no. 12 (December
  2008): 1814--21, \url{https://doi.org/10.1345/aph.1L474}; Hwang,
  Thomas J., Florence T. Bourgeois, and John D. Seeger. ``Drug Safety in
  the Digital Age.'' \emph{New England Journal of Medicine} 370, no. 26
  (June 26, 2014): 2460--62. \url{https://doi.org/10.1056/NEJMp1401767};
  Jona Kräenbring et al., ``Accuracy and Completeness of Drug
  Information in Wikipedia: A Comparison with Standard Textbooks of
  Pharmacology,'' \emph{PLOS ONE} 9, no. 9 (September 24, 2014):
  e106930, \url{https://doi.org/10.1371/journal.pone.0106930}; N. J.
  Reavley , A. J. Mackinnon , A. J. Morgan , M. Alvarez-Jimenez , S. E.
  Hetrick , E. Killackey , B. Nelson , R. Purcell , M. B. H. Yap, and A.
  F. Jorm, ``Quality of Information Sources about Mental Disorders: A
  Comparison of Wikipedia with Centrally Controlled Web and Printed
  Sources,'' \emph{Psychological Medicine} 42, no. 8 (August 2012):
  1753--62, \url{https://doi.org/10.1017/S003329171100287X}; Peter G.
  Volsky, Cristina M. Baldassari, Sirisha Mushti, and Craig S. Derkay,
  ``Quality of Internet Information in Pediatric Otolaryngology: A
  Comparison of Three Most Referenced Websites,'' \emph{International
  Journal of Pediatric Otorhinolaryngology} 76, no. 9 (September 2012):
  1312--6,
  \url{https://doi.org/10.1016/j.ijporl.2012.05.026}.
\item
  Shane Greenstein and Feng Zhu, ``Is Wikipedia Biased?'' \emph{American
  Economic Review} 102, no. 3 (May 2012): 343--48,
  \url{https://doi.org/10.1257/aer.102.3.343}; Shane Greenstein and Feng
  Zhu, ``Do Experts or Crowd-Based Models Produce More Bias? Evidence
  from Encyclopedia Britannica and Wikipedia,'' \emph{Management
  Information Systems Quarterly} 42, no. 3 (September 1, 2018): 945--59,
  \url{https://aisel.aisnet.org/misq/vol42/iss3/14}.
\item
  Morten Warncke-Wang, Vivek Ranjan, Loren Terveen, and Brent Hecht.
  ``Misalignment Between Supply and Demand of Quality Content in Peer
  Production Communities,'' in \emph{Proceedings of the Ninth
  International AAAI Conference on Web and Social Media (ICWSM '15)}
  (AAAI, 2015), 493--502,
  \url{https://www.aaai.org/ocs/index.php/ICWSM/ICWSM15/paper/viewFile/10591/10532}.
\item
  Aniket Kittur and Robert E. Kraut, ``Harnessing the Wisdom of Crowds
  in Wikipedia: Quality Through Coordination,'' in \emph{Proceedings of
  the 2008 ACM Conference on Computer Supported Cooperative Work (CSCW
  '08)} (New York, NY: ACM, 2008), 37--46,
  \url{https://doi.org/10.1145/1460563.1460572}; Warncke-Wang et al.,
  ``Misalignment Between Supply and Demand of Quality Content in Peer
  Production
  Communities.''
\item
  Feng Shi, Misha Teplitskiy, Eamon Duede, and James A. Evans, ``The
  Wisdom of Polarized Crowds,'' \emph{Nature Human Behaviour} 3, no. 4
  (April 2019): 329,
  \url{https://doi.org/10.1038/s41562-019-0541-6}.
\item
  W. Ben Towne, Aniket Kittur, Peter Kinnaird, and James Herblseb,
  ``Your Process Is Showing: Controversy Management and Perceived
  Quality in Wikipedia,'' in \emph{Proceedings of the 2013 Conference on
  Computer Supported Cooperative Work (CSCW '13)} (New York, NY: ACM,
  2013), 1059--68,
  \url{https://doi.org/10.1145/2441776.2441896}.
\item
  Hsin-liang Chen, ``The Perspectives of Higher Education Faculty on
  Wikipedia,'' \emph{The Electronic Library} 28, no. 3 (June 8, 2010):
  361--73, \url{https://doi.org/10.1108/02640471011051954}; Aline
  Soules, ``Faculty Perception of Wikipedia in the California State
  University System,'' \emph{New Library World} 116, no. 3/4 (March 9,
  2015): 213--26,
  \url{https://doi.org/10.1108/NLW-08-2014-0096}.
\item
  Christina Shane-Simpson, Elizabeth Che, and Patricia J. Brooks,
  ``Giving Psychology Away: Implementation of Wikipedia Editing in an
  Introductory Human Development Course,'' \emph{Psychology Learning \&
  Teaching} 15, no. 3 (November 1, 2016): 268--93,
  \url{https://doi.org/10.1177/1475725716653081}.
\item
  Jonathon Penney, ``Chilling Effects: Online Surveillance and Wikipedia
  Use,'' \emph{Berkeley Technology Law Journal} 31, no. 1 (June 1,
  2016): 117,
  \url{https://doi.org/10.15779/Z38SS13}.
\item
  David J. McIver and John S. Brownstein, ``Wikipedia Usage Estimates
  Prevalence of Influenza-Like Illness in the United States in Near
  Real-Time,'' \emph{PLoS Computational Biology} 10, no. 4 (April 17,
  2014): e1003581, \url{https://doi.org/10.1371/journal.pcbi.1003581};
  Márton Mestyán, Taha Yasseri, and János Kertész, ``Early Prediction of
  Movie Box Office Success Based on Wikipedia Activity Big Data,''
  \emph{PLOS ONE} 8, no. 8 (August 21, 2013): e71226,
  \url{https://doi.org/10.1371/journal.pone.0071226}; Taha Yasseri and
  Jonathan Bright, ``Wikipedia Traffic Data and Electoral Prediction:
  Towards Theoretically Informed Models,'' \emph{EPJ Data Science} 5,
  no. 1 (June 18, 2016): 22,
  \url{https://doi.org/10.1140/epjds/s13688-016-0083-3}; Benjamin K.
  Smith and Abel Gustafson, ``Using Wikipedia to Predict Election
  Outcomesonline Behavior as a Predictor of Voting,'' \emph{Public
  Opinion Quarterly} 81, no. 3 (September 7, 2017): 714--35,
  \url{https://doi.org/10.1093/poq/nfx007}; Gabriele Tolomei et al.,
  ``Twitter Anticipates Bursts of Requests for Wikipedia Articles,'' in
  \emph{Proceedings of the 2013 Workshop on Data-Driven User Behavioral
  Modelling and Mining from Social Media}, DUBMOD '13 (New York, NY:
  ACM, 2013), 5--8, \url{https://doi.org/10.1145/2513577.2538768};
  Ferron, Michela, and Paolo Massa. ``Beyond the Encyclopedia:
  Collective Memories in Wikipedia.'' \emph{Memory Studies} 7, no. 1
  (January 1, 2014): 22--45.
  \url{https://doi.org/10.1177/1750698013490590}.
\item
  Reid Priedhorsky, Jilin Chen, Shyong (Tony) K. Lam, Katherine
  Panciera, Loren Terveen, John Riedl, ``Creating, Destroying, and
  Restoring Value in Wikipedia,'' in \emph{Proceedings of the 2007
  International ACM Conference on Supporting Group Work}, GROUP '07 (New
  York, NY: ACM, 2007), 259--68,
  \url{https://doi.org/10.1145/1316624.1316663}.
\item
  Andreea D. Gorbatâi, ``The Paradox of Novice Contributions to
  Collective Production: Evidence from Wikipedia,'' SSRN Scholarly Paper
  (Rochester, NY: Social Science Research Network, February 10, 2014),
  \url{https://papers.ssrn.com/abstract=1949327}.
\item
  e.g., David A. Hoffman and Salil K. Mehra, ``Wikitruth Through
  Wikiorder,'' \emph{Emory Law Journal}, no. 1 (2009--2010): 151--210,
  \url{https://heinonline.org/HOL/P?h=hein.journals/emlj59\&i=153};
  Piotr Konieczny, ``Governance, Organization, and Democracy on the
  Internet: The Iron Law and the Evolution of Wikipedia,''
  \emph{Sociological Forum} 24, no. 1 (2009): 162--92,
  \url{https://doi.org/10.1111/j.1573-7861.2008.01090.x}; Piotr
  Konieczny, ``Adhocratic Governance in the Internet Age: A Case of
  Wikipedia,'' \emph{Journal of Information Technology \& Politics} 7,
  no. 4 (October 11, 2010): 263--83,
  \url{https://doi.org/10.1080/19331681.2010.489408}; Joseph Reagle,
  \emph{Good Faith Collaboration: The Culture of Wikipedia} (Cambridge,
  MA: MIT Press, 2010); Dariusz Jemielniak, \emph{Common Knowledge?: An
  Ethnography of Wikipedia} (Stanford, CA: Stanford University, 2014);
  Emiel Rijshouwer, \emph{Organizing Democracy: Power Concentration and
  Self-Organization in the Evolution of Wikipedia} (Rotterdam: Erasmus
  University, 2019),
  \url{https://repub.eur.nl/pub/113937/}.
\item
  Andrea Forte, Vanesa Larco, and Amy Bruckman, ``Decentralization in
  Wikipedia Governance,'' \emph{Journal of Management Information
  Systems} 26, no. 1 (July 1, 2009): 49--72,
  \url{https://doi.org/10.2753/MIS0742-1222260103}; Loxley Sijia Wang,
  Jilin Chen, Yuqing Ren, and John Reidl, ``Searching for the Goldilocks
  Zone: Trade-Offs in Managing Online Volunteer Groups,'' in
  \emph{Proceedings of the ACM 2012 Conference on Computer Supported
  Cooperative Work}, CSCW '12 (New York, NY, USA: ACM, 2012), 989--98,
  \url{https://doi.org/10.1145/2145204.2145351}; Haiyi Zhu, Robert E.
  Kraut, and Aniket Kittur, ``Organizing Without Formal Organization:
  Group Identification, Goal Setting and Social Modeling in Directing
  Online Production,'' in \emph{Proceedings of the ACM 2012 Conference
  on Computer Supported Cooperative Work}, CSCW '12 (New York, NY, USA:
  ACM, 2012), 935--44, \url{https://doi.org/10.1145/2145204.2145344};
  Jonathan T. Morgan, Michael Gilbert, David W. McDonald, and Mark
  Zachry, ``Project Talk: Coordination Work and Group Membership in
  WikiProjects,'' in \emph{Proceedings of the 9th International
  Symposium on Open Collaboration}, WikiSym '13 (New York, NY, USA: ACM,
  2013), 3:1--3:10,
  \url{https://doi.org/10.1145/2491055.2491058}.
\item
  Aaron Shaw and Benjamin Mako Hill, ``Laboratories of Oligarchy? How
  the Iron Law Extends to Peer Production,'' \emph{Journal of
  Communication} 64, no. 2 (2014): 215--38,
  \url{https://doi.org/10.1111/jcom.12082}.
\item
  Benjamin Mako Hill and Aaron Shaw, ``Studying Populations of Online
  Communities,'' in \emph{The Oxford Handbook of Networked
  Communication}, ed. Brooke Foucault Welles and Sandra González-Bailón,
  (Oxford, UK: Oxford University Press, 2019),
  \url{https://doi.org/10.1093/oxfordhb/9780190460518.013.8}.
\item
  Aaron Halfaker, R. Stuart Geiger, Jonathan T. Morgan, and John Riedl,
  ``The Rise and Decline of an Open Collaboration System: How
  Wikipedia's Reaction to Popularity Is Causing Its Decline,''
  \emph{American Behavioral Scientist} 57, no. 5 (May 1, 2013): 664--88,
  \url{https://doi.org/10.1177/0002764212469365}; Shaw and Hill,
  ``Laboratories of Oligarchy?''; Nathan TeBlunthuis, Aaron Shaw, and
  Benjamin Mako Hill, ``Revisiting `The Rise and Decline' in a
  Population of Peer Production Projects,'' in \emph{Proceedings of the
  2018 CHI Conference on Human Factors in Computing Systems (CHI '18)}
  (New York, NY: ACM, 2018), 355:1--355:7,
  \url{https://doi.org/10.1145/3173574.3173929}.
\item
  e.g., Felipe Ortega, ``Wikipedia: A Quantitative Analysis'' (Ph.D.
  dissertation, Universidad Rey Juan Carlos, 2009),
  \url{http://libresoft.es/Members/jfelipe/phd-thesis}.
\item
  Sneha Narayan, Jake Orlowitz, Jonathan T. Morgan, Benjamin Mako Hill,
  and Aaron Shaw. ``The Wikipedia Adventure: Field Evaluation of an
  Interactive Tutorial for New Users,'' in \emph{Proceedings of the 2017
  ACM Conference on Computer Supported Cooperative Work and Social
  Computing}, CSCW '17 (New York, NY, USA: ACM, 2017), 1785--99,
  \url{https://doi.org/10.1145/2998181.2998307}.
\item
  Morgan Michelle Stoddard, ``Judicial Citation to Wikipedia in
  Published Federal Court'' (Masters in Library Science, UNC, 2009),
  \url{http://ils.unc.edu/MSpapers/3526.pdf}.
\item
  Nicholas Vincent, Isaac Johnson, and Brent Hecht, ``Examining
  Wikipedia with a Broader Lens: Quantifying the Value of Wikipedia's
  Relationships with Other Large-Scale Online Communities,'' in
  \emph{Proceedings of the 2018 CHI Conference on Human Factors in
  Computing Systems}, CHI '18 (New York, NY: ACM, 2018), 566:1--566:13,
  \url{https://doi.org/10.1145/3173574.3174140}; Connor McMahon, Isaac
  L. Johnson, and Brent J. Hecht, ``The Substantial Interdependence of
  Wikipedia and Google: A Case Study on the Relationship Between Peer
  Production Communities and Information Technologies,'' in
  \emph{International AAAI Conference on Web and Social Media (ICWSM
  2017)} (Palo Alto, California: AAAI, 2017), 142--51,
  \url{http://brenthecht.com/publications/icwsm17_googlewikipedia.pdf}.
\item
  Marit Hinnosaar et al., ``Wikipedia Matters,'' SSRN Scholarly Paper
  (Rochester, NY: Social Science Research Network, July 14, 2019),
  \url{https://papers.ssrn.com/abstract=3046400}.
\item
  Neil Thompson and Douglas Hanley, ``Science Is Shaped by Wikipedia:
  Evidence from a Randomized Control Trial,'' SSRN Scholarly Paper
  (Rochester, NY: Social Science Research Network, February 13, 2018),
  \url{https://papers.ssrn.com/abstract=3039505}; Mark Zastrow,
  ``Wikipedia Shapes Language in Science Papers,'' \emph{Nature News},
  September 26, 2017,
  \url{https://doi.org/10.1038/nature.2017.22656}.
\end{enumerate}

\end{document}